\documentclass[twocolumn,tighten]{aastex63}

\usepackage{graphicx}
\usepackage{amsmath}
\usepackage{amssymb}
\usepackage{color}
\usepackage{mathtools}
\usepackage{ulem}
\usepackage[all]{hypcap} 

\usepackage{lineno}

\newcommand\msun{\, {M}_\odot}

\newcommand\gpcyr{\, {\rm Gpc}^{-3}\,{\rm yr}^{-1}}

\def\apjl{ApJL}%
\def\apjs{ApJS}%
\def\aap{A\&A}%
\begin{document}

\title{Constraining the cosmic merger history of intermediate-mass black holes with gravitational wave detectors}
\shorttitle{Constraining the merger history of IMBHs}
\correspondingauthor{Giacomo Fragione}
\email{giacomo.fragione@northwestern.edu}

\author[0000-0002-7330-027X]{Giacomo Fragione}
\affil{Center for Interdisciplinary Exploration \& Research in Astrophysics (CIERA), Northwestern University, Evanston, IL 60208, USA}
\affil{Department of Physics \& Astronomy, Northwestern University, Evanston, IL 60208, USA}

\author[0000-0003-4330-287X]{Abraham Loeb}
\affil{Astronomy Department, Harvard University, 60 Garden St., Cambridge, MA 02138, USA}

\begin{abstract}
Intermediate-mass black holes (IMBHs) have not been detected beyond any reasonable doubt through either dynamical or accretion signatures. Gravitational waves (GWs) represent an unparalleled opportunity to survey the sky and detect mergers of IMBHs, making it possible for the first time to constrain their formation, growth, and merger history across cosmic time. While the current network LIGO-Virgo-KAGRA is significantly limited in detecting mergers of IMBH binaries, the next generation of ground-based observatories and space-based missions promise to shed light on the IMBH population through the detection of several events per year. Here, we asses this possibility by determining the optimal network of next-generation of GW observatories to reconstruct the IMBH merger history across cosmic time. We show that Voyager, the Einstein Telescope, and Cosmic Explorer will be able to constrain the distribution of the primary masses of merging IMBHs up to $\sim 10^3\msun$ and with mass ratio $\gtrsim 0.1$, while LISA will complementary do so at higher mass and smaller mass ratios. Therefore, a network of next-generation ground-based and space-based observatories will potentially reconstruct the merger history of IMBHs. Moreover, IMBHs with masses $\lesssim 5\times 10^3\ M_\odot$ could be observed in multiband up to a redshift of $z\approx 4$, ushering in a new of era GW astronomy.
\end{abstract}

\section{Introduction}
\label{sect:intro}

Intermediate-mass black holes (IMBHs) are the missing link between stellar-mass BHs, with masses $\lesssim 100\msun$, and supermassive BHs, with masses $\gtrsim 10^5\msun$ \citep*[see][for a review]{GreeneStrader2020}. While there is plenty of evidence for the existence of the other two subfamilies, IMBHs have not been confirmed beyond any reasonable doubt. Nevertheless, IMBHs could play a fundamental role in the evolution of galaxies, as they could be the seeds that later become supermassive BHs through mergers and accretion \citep[e.g.,][]{madau2001,Silk2017,Natarajan2021}, and they can be source of tidal disruption events \citep[e.g.,][]{RosswogRamirez-Ruiz2009,MacLeodGuillochon2016}, ultra-luminous X-ray binaries \citep[e.g.,][]{kaaret2017ARA&A..55..303K} and gravitational waves \citep[GWs; e.g.,][]{MandelBrown2008,gair2011,fragl2018b}. 

There exist three main formation channels for IMBHs. Direct collapse of a gas cloud of pristine gas \citep{bromm2003,begelm2006} or the evolution of massive Pop III stars \citep{madau2001,bromm2004} take place at high redshift, and might form an IMBH of $\sim 10^4 - 10^5\msun$ and $\sim 100\msun$, respectively. Repeated mergers either of massive main-sequence stars, later collapsing to form a BH, \citep{por02,gurk2004,gie15,DiCarloMapelli2021,GonzalezKremer2021}, or of stellar-mass BHs \citep{mil02b,antonini2019,FragioneKocsis2022} could produce IMBHs with masses in the range $\sim 10^2 - 10^4\msun$. Other channels may possibly include super-Eddington accretion onto stellar BHs embedded in the disks of active galactic nuclei \citep[e.g.,][]{koc11} and repeated collisions of BHs with stars \citep[e.g.,][]{StoneKupper2017,RizzutoNaab2022,RoseNaoz2022}. 

Since we observe the two mass brackets of the BH population, IMBHs must have existed at some point across cosmic time. Based on their phenomenology, there are four main observational strategies to detect them. Tracking the motion of stars and gas and constraining accretion have been proven to be able to detect massive candidates ($\sim 10^4 - 10^5\msun$), but may suffer from several systematics and are realistically limited to IMBHs within $\sim 10-100$ Mpc \citep[e.g.,][]{BaldassareReines2015,chili2018,PechettiSeth2022}. Looking for transient events, such as tidal disruption events (in particular of white dwarfs), in the outskirts of a galaxy is a promising method for moderately-massive candidates out to redshift $z\lesssim 1$, as possibly already done in a few instances \citep[e.g.,][]{lin2018,peng2019,shen2019}. The most promising way to find IMBHs is looking for GW events, where one of the components in the merging binary is in the $[10^2-10^4]\msun$ mass range. While LIGO/Virgo/KAGRA could detect the merger of IMBHs with masses $\sim 100\,M_\odot$ out to $z\sim 1$ \citep[e.g.,][]{AbbottAbbott2019}, the upcoming space-based LISA and ground-based Einstein Telescope (ET) and Cosmic Explorer (CE) offer an unparalleled opportunity of detecting IMBHs up to $z\sim10-100$ \citep[e.g.,][]{Amaro-SeoaneAudley2017,JaniShoemaker2020}.

Only a few studies in the literature have systematically computed the merger rate of IMBH binaries in various astrophysical scenarios, and the characteristics of the observable merging population for present and upcoming observatories. Within the large uncertainties of the astrophysical models, merger rates of IMBHs are typically predicted to be in the broad range $\sim 10^{-3}-10\gpcyr$ in the local Universe, typically with a peak in redshift from $z\approx 1$ to $z\approx 5$, implying several detectable IMBH mergers per year with upcoming GW observatories \citep[e.g.,][]{Miller2002,MandelBrown2008,fragk18,fragl2018b,fragleiginkoc18,agu18,antonini2019,Arca-SeddaCapuzzo-Dolcetta2019,RasskazovFragione2020,ArcaSeddaAmaroSeoane2021,DiMatteoNi2022,FragioneKocsis2022,HijikawaKinugawa2022}. Therefore, it is critical to answer the question whether, given an observed population of merging IMBH binaries, the detected systems tracks the properties of the underlying astrophysical population. To answer this question, we assume a rather universal distribution of mergers of IMBH binaries, and show that a combination of next-generation ground-based and space-based observatories could constrain the cosmic IMBH merger history.

This paper is organized as follows. In Section~\ref{sect:method}, we describe our model and how we sample a population of merging IMBH binaries across redshift, primary masses, and mass ratio. In Section~\ref{sect:results}, we determine the fraction of observable events, and illustrate how the detected population of merging IMBH binaries by ground-based and space-based detectors can be used to map the properties of the underlying astrophysical population. Finally, in Section~\ref{sect:concl}, we discuss the implications of our results and draw our conclusions.

\section{Merging binaries}
\label{sect:method}

In what follows, we describe the details of the method we use to sample a population of merging IMBH binaries across redshift, primary masses, and mass ratios.

We assume that the volumetric merger rate of IMBHs is described by a function
\begin{equation}
    R(z,M_1,q) = K \mathcal{N}(\mu_z, \sigma_z) M_1^{-\alpha} q^{-\beta}\,,
    \label{eqn:rate}
\end{equation}
where $K$ is a normalization constant (with the units of Gpc$^{-3}$ yr$^{-1}$), $\mathcal{N}(\mu_z, \sigma_z)$ is the redshift-dependent part of the merger rate, which we model for simplicity as a normal distribution with mean $\mu_z$ and standard deviation $\sigma_z$ reminiscent of the trends found in literature \citep[e.g.,][]{fragk18,fragleiginkoc18,RasskazovFragione2020,Fragione2022,FragioneKocsis2022,HijikawaKinugawa2022}, and $\alpha$ and $\beta$ are the exponents of the primary mass ($M_1$, in units of Solar masses) and mass-ratio ($q$) distributions, respectively, assumed to be described by power laws. Therefore, the number of events per year, primary mass, and mass ratio will simply be
\begin{eqnarray}
    \frac{d^3 \dot{N}}{dz dM_1 dq} &=& \frac{R(z,M_1,q)}{1+z} \frac{dV_c}{dz} = \nonumber\\
    &=&  K\frac{\mathcal{N}(\mu_z, \sigma_z)}{1+z} \frac{dV_c}{dz} M_1^{-\alpha} q^{-\beta}\,,
    \label{eqn:nevents}
\end{eqnarray}
where $dV_c/dz$ is the differential comoving volume; the number of detectable events will be given by
\begin{equation}
    \frac{d^3 \dot{N}_{\rm det}}{dz dM_1 dq} = F_{\rm det}(z,M_1,q) \frac{d^3 \dot{N}}{dz dM_1 dq}\,,
\end{equation}
where $F_{\rm det}(z,M_1,q)$ is the detector-dependent detectability function, which encodes the ability of observing the merger of a binary with primary mass $M_1$ and mass ratio $q$ at a redshift $z$. For simplicity, we model this function as
\begin{equation}
    F_{\rm det}(z,M_1,q) = H\left(\left\langle \rho (z,M_1,q) \right\rangle > \rho_{\rm thr}\right)\,,
    \label{eqn:fdet}
\end{equation}
where $H$ is the Heaviside function, $\left\langle \rho (z,M_1,q) \right\rangle$ is the averaged (over sky locations) signal-to-noise (SNR) ratio, and $\rho_{\rm thr}=8$ is the threshold SNR for a detection.

The average SNR of a merging system in a detector frequency band is computed as
\begin{equation}
\left\langle \rho (z,M_1,q) \right\rangle=2\mathcal{C} \sqrt{\int_{f_{\rm min}}^{f_{\rm max}} \frac{|\tilde{h}(f)|^2}{S_n(f)} df}\,,
\label{eqn:rhof}
\end{equation}
where $\mathcal{C}=2/\sqrt{5}$ and $\mathcal{C}=2/5$ for space-based and ground-based detectors (obtained from averaging over sky locations), respectively \citep{robson2019}. In the previous equation, $f_{\rm min}$ and $f_{\rm max}$ are the minimum and maximum frequency of the binary in the dectector band, respectively, $S_n(f)$ is the noise power spectral density, and $|\tilde{h}(f)|$ is the frequency-domain waveform amplitude for a face-on binary.  We use \textsc{pyCBC} \citep{NitzHarry2019} with the IMRPhenomD approximant \citep{HusaKhan2016} to compute the waveform of merging IMBH binaries (assuming non-spinning binaries). Note that the strength of the GW signal is proportional to $\mathcal{M}_\mathrm{c,z}^{5/6}$, where the redshifted chirp mass is defined as
\begin{equation}
\mathcal{M}_\mathrm{c,z}=(1+z) M_1\frac{q^{3/5}}{(1+q)^{1/5}}\,,
\label{eqn:mchirpz}
\end{equation}
and inversely proportional to the luminosity distance of the event. In our calculations, we assume the standard $\Lambda$CDM cosmology \citep{PlanckCollaborationAde2016}.

\begin{figure*} 
\centering
\includegraphics[scale=0.525]{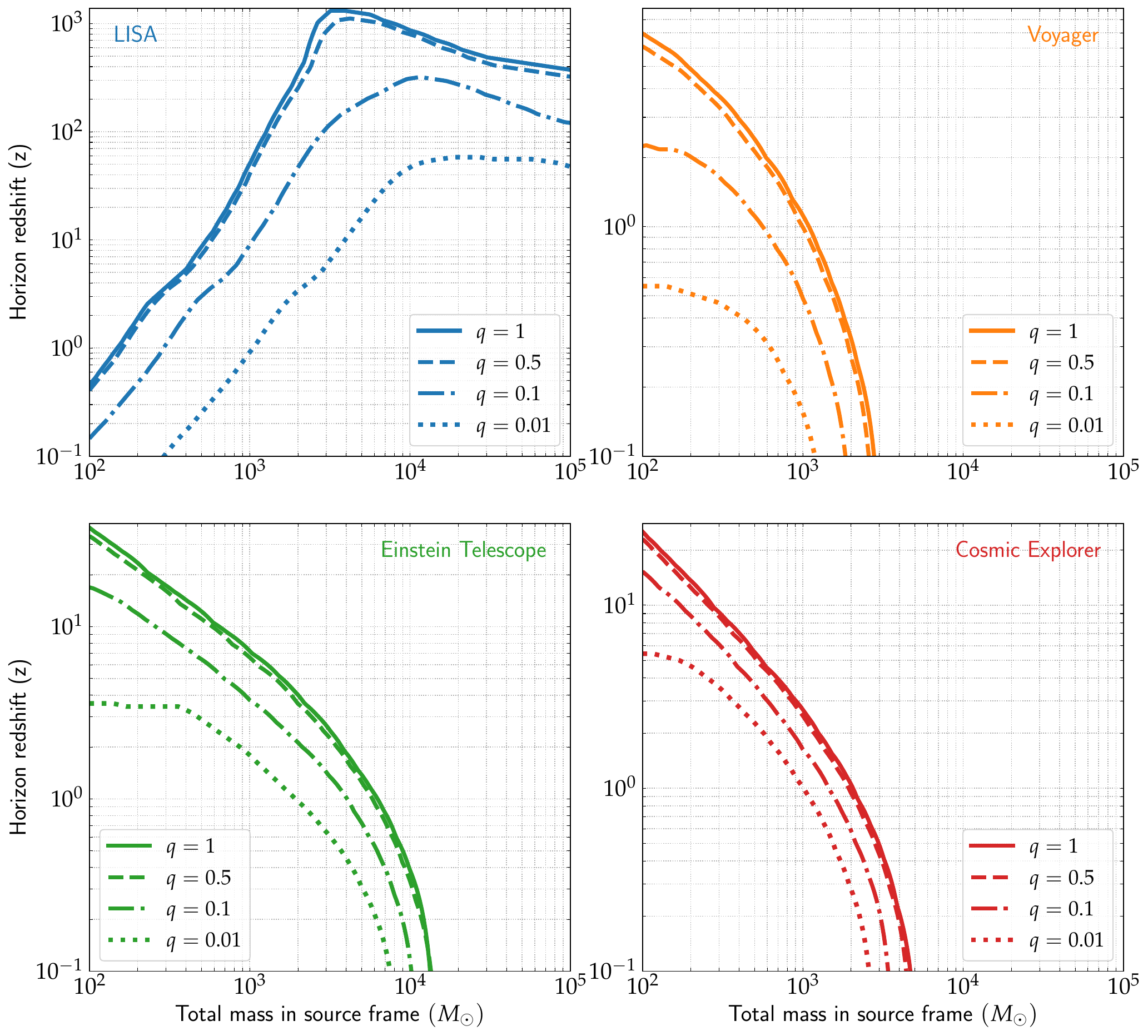}
\caption{Horizon distance (measured as the cosmological redshift for $\Lambda$CDM cosmology) at the minimum threshold (SNR of $8$) for IMBH binaries with different mass ratios. Top-left: LISA; top-right: Voyager; bottom-left: Einstein Telescope; bottom-right: Cosmic Explorer.}
\label{fig:sens}
\end{figure*}

\begin{figure*} 
\centering
\includegraphics[scale=0.675]{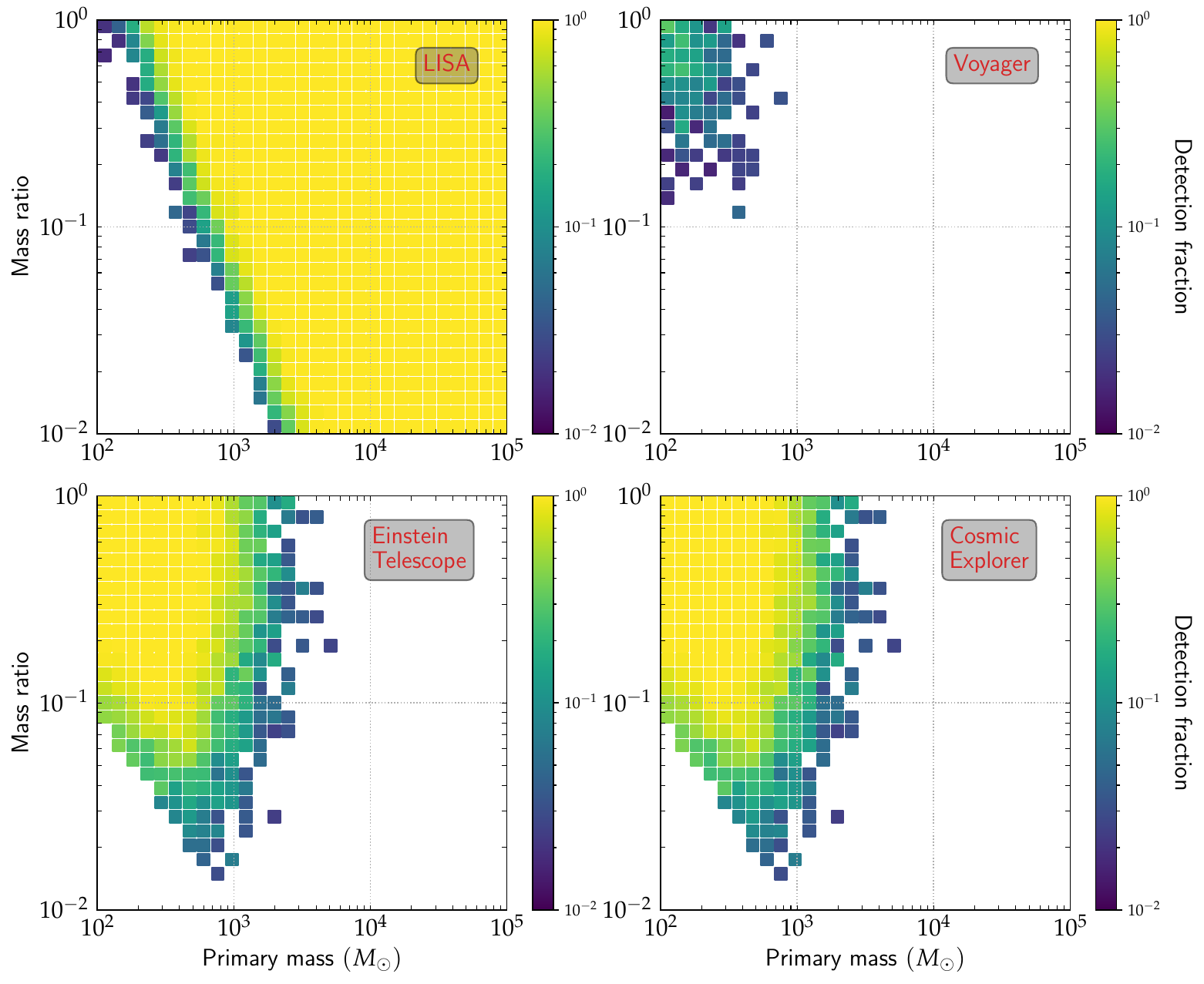}
\caption{Detection fraction for IMBH binaries as a function of the primary mass and mass ratio. Top left: LISA; top right: Voyager; bottom left: Einstein Telescope; bottom right: Cosmic Explorer.}
\label{fig:instr}
\end{figure*}

\section{The detectable population}
\label{sect:results}

\begin{figure} 
\centering
\includegraphics[scale=0.55]{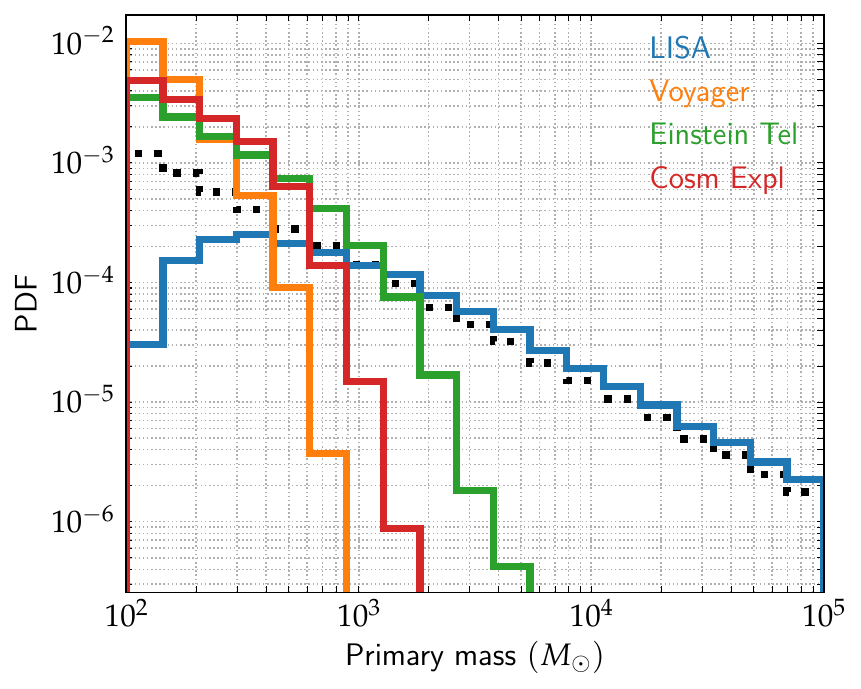}
\includegraphics[scale=0.55]{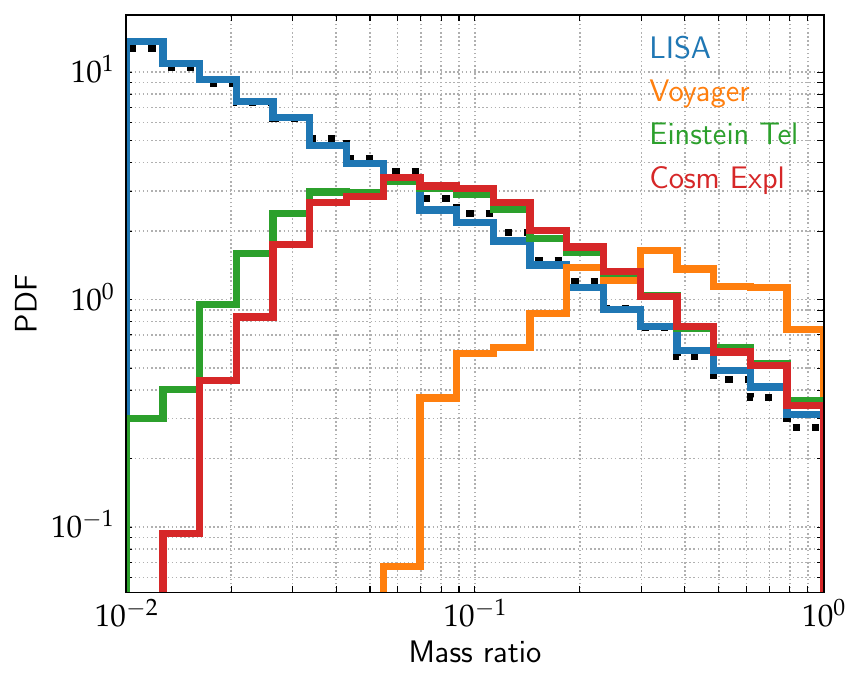}
\includegraphics[scale=0.55]{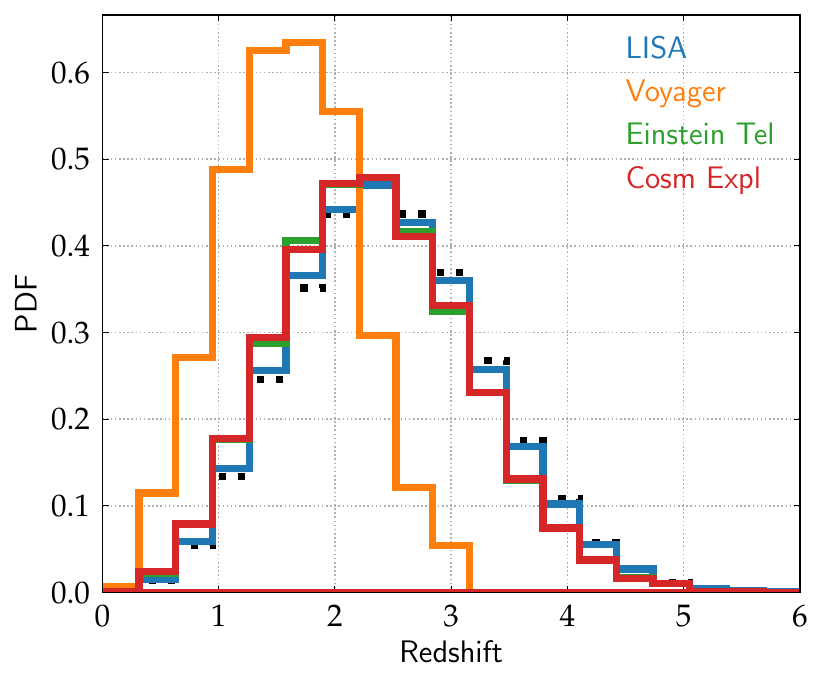}
\caption{Probability distribution function of primary mass (top), mass ratio (center), and redshift (bottom) of merging IMBH binaries detected by LISA (blue), Voyager (orange), Einstein Telescope (green), and Cosmic Explorer (red). The dotted black line represents the underlying sampled population from Eq.~\ref{eqn:nevents} when $\{\mu_z,\sigma_z,\alpha,\beta\}=\{2,1,1,1\}$.}
\label{fig:muz2}
\end{figure}

\begin{figure} 
\centering
\includegraphics[scale=0.55]{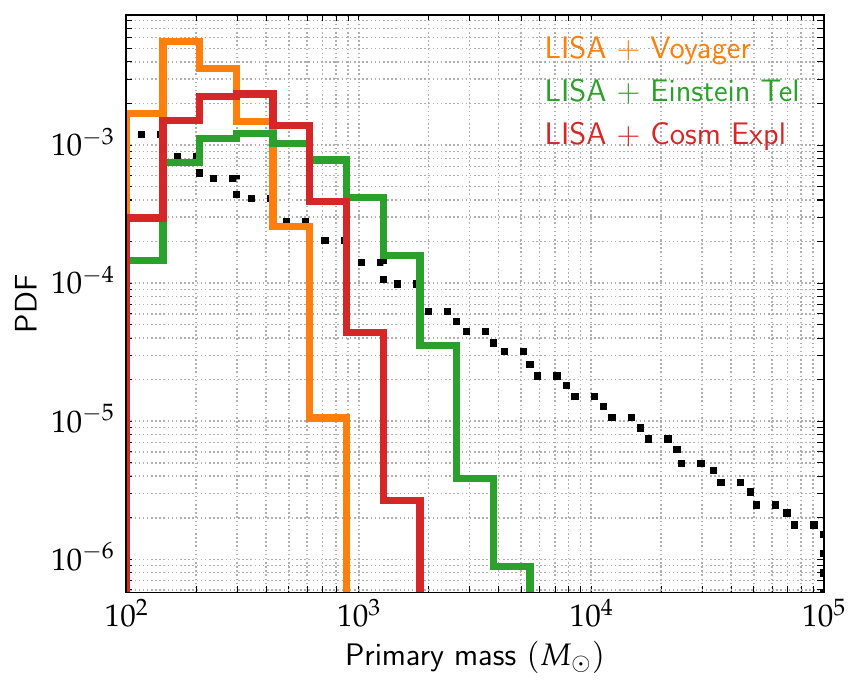}
\includegraphics[scale=0.55]{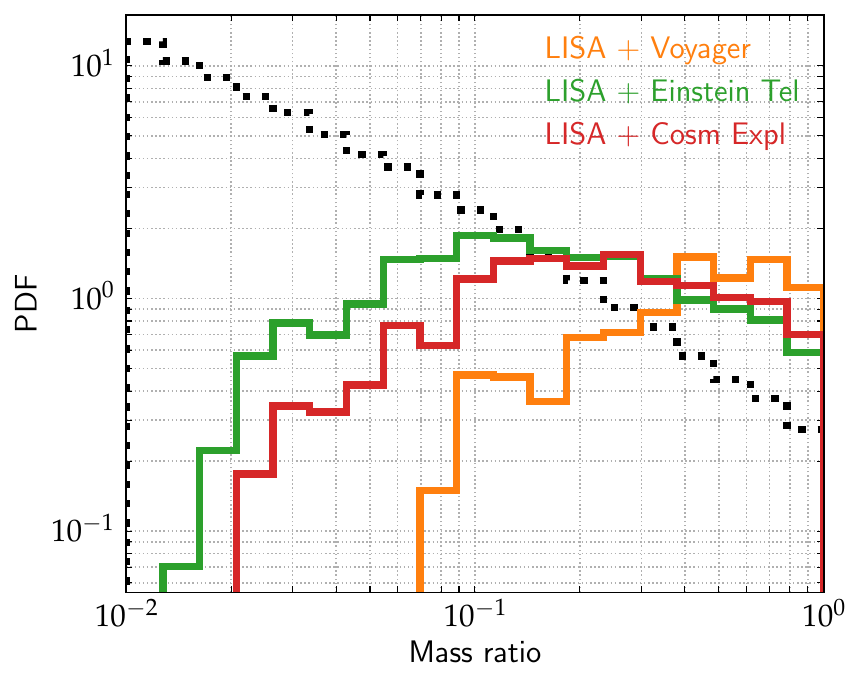}
\includegraphics[scale=0.55]{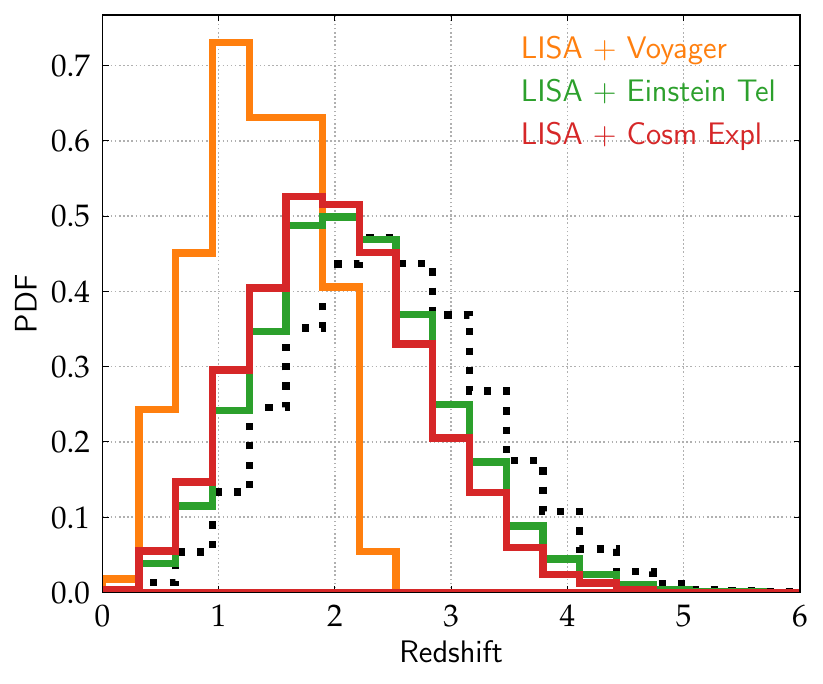}
\caption{Probability distribution function of primary mass (top), mass ratio (center), and redshift (bottom) of merging IMBH binaries detected in multi-band by LISA+Voyager (orange), LISA + Einstein Telescope (green), and LISA + Cosmic Explorer (blue). The dotted black line represents the underlying sampled population from Eq.~\ref{eqn:nevents} when $\{\mu_z,\sigma_z,\alpha,\beta\}=\{2,1,1,1\}$.}
\label{fig:muz2_multi}
\end{figure}

In what follows, we present the results of our analysis and discuss how the merging binary population detected by a combination of ground-based and space-based observatories could constrain the underlying IMBH merger history. For a given choice of $\{\mu_z,\sigma_z,\alpha,\beta\}$, we sample $10^5$ events from Eq.~\ref{eqn:nevents} using our publicly-available code \textsc{imbhistory} \citep{imbhistory}. The primary mass is drawn in the range $[10^2-10^5]\msun$, while the mass ratio distribution in the range $[0.01,1]$. Note that we discard every system whose secondary is $<10\msun$.

We consider four different GW observatories, namely LISA, Voyager (the upgraded detector of the current LIGO facility), ET, and CE. We compute the power spectral density of LISA as in \citet{robson2019}, of Voyager as in \citet{LIGOVoy2018}, of ET as in \citet{PunturoAbernathy2010}, and of CE as in \citet{ReitzeAdhikari2019}. The space-based mission LISA has a planned of $T_{\rm LISA}=5$ yr. In this interval of time, a binary will evolve towards a merger starting from the frequency \citep[see Eq.~24 in][]{robson2019}
\begin{eqnarray}
    f_{\rm ini} &=& 1.2\times 10^{-2}\,{\rm Hz}\ (1+z)^{-5/8} \left(\frac{1+q}{q^3}\right)^{1/8}\nonumber\\
    & \times &\left(\frac{m_1}{100\msun}\right)^{-5/8} \left(\frac{T_{\rm LISA}}{4\,{\rm yr}}\right)^{-3/8}\,.
\end{eqnarray}
This frequency will typically be higher than the minimum detectable frequency for LISA, nominally $10^{-5}$\,Hz. Therefore, $f_{\rm min}=f_{\rm ini}$ in Eq.~\ref{eqn:rhof} for LISA. For ground-based detectors, this frequency is always smaller than the frequency at which Voyager, ET, and CE start operating, $5$\,Hz, $1$\,Hz, and $5$\,Hz, respectively.

Figure~\ref{fig:sens} illustrates the horizon distance (measured as the cosmological redshift for $\Lambda$CDM cosmology) at the minimum threshold SNR ($\rho_{\rm thr}=8$) for IMBH binaries with different mass ratios. Voyager, ET, and CE are most sensitive to lower-mass binaries; their horizon distances for equal-mass binaries of $\sim 100\msun$ are of about $7$, $40$, $30$, respectively, which then decrease for higher binary masses. Indeed, these observatories are mostly sensitive to the merger and ringdown phases of IMBH binaries, with possibly just a few inspiral cycles within their sensitivity band. In contrast, LISA performs poorly in detecting these low-mass binaries since they only spend a small fraction of their early inspiral in its detection band. However, LISA is able to detect equal-mass IMBH binaries of $\sim 10^3\msun$ up to a redshift of $\sim 10^3$, which typically spend several year in the LISA frequency band before merging. For non-equal-mass binaries, the horizon distance decrease at a given total binary mass, as a consequence of the fact that the chirp mass becomes smaller for smaller mass ratios (see Eq.~\ref{eqn:mchirpz}). While $q=0.5$ does not affect significantly the horizon distance of any of the observatories, we find that the horizon distance decreases by about an order of magnitude with respect to the case of equal-mass binaries when $q=0.01$.

Figure~\ref{fig:instr} illustrates the detection fraction, i.e. the number of detectable binaries out of the total sampled population as a function of the primary mass and mass ratio for $\{\mu_z,\sigma_z,\alpha,\beta\}=\{2,1,1,1\}$. For example, this choice of parameters could represent the merger history of IMBH binaries catalyzed from repeated mergers in dense star clusters \citep[e.g.,][]{FragioneKocsis2022}. We find that LISA can detect any merging system with primary mass $\gtrsim 10^3\msun$ for any mass ratio, while for primary masses $\sim 10^3\msun$ only when $q>0.1$. At smaller masses, LISA can only observe a few percent of the merging IMBH binaries, as a result that only spend a small fraction of their early inspiral in its detection band. Voyager could find only about $10\%$ of near equal-mass systems with component masses in the range $[100\msun-400\msun]$, while it is not sensitive enough to observe the mergers of $\sim 10^3\msun$ IMBHs. Finally, ET and CE are able to detect any binary with $M_1<10^3\msun$ for $q>0.1$, while about $10\%$ of the systems at smaller mass ratios or with larger primary masses (up to about $2000\msun$).

To further break down the detection landscape of upcoming ground-based and space-based detectors, we plot in Figure~\ref{fig:muz2} the probability distribution function of the primary mass (top) of merging IMBH binaries. It is clear that LISA does an exquisite job in reproducing the shape of the distribution of primary masses of merging binaries for $M_1\gtrsim 10^3\msun$. This is evidenced by the fact that the detected population closely follows the true underlying astrophysical one (black dotted line). At lower masses, the systems observed by LISA do not reproduce the characteristic slope of the primary mass distribution. As said, LISA can only detect a few percent of the events in this part of the parameter space since these binaries will only spend a few cycles in its detection band, before exiting it on their way to a merger. The characteristics of the astrophysical population of merging IMBH binaries at $M_1<10^3\msun$ can be recovered with a good approximation by using ground-spaced observatories. Voyager, ET, and CE are able to reproduce the underlying distribution for primary masses up to about $400\msun$, $2\times 10^3\msun$, and $10^3\msun$, respectively. 

In the central panel of Figure~\ref{fig:muz2}, we show the probability distribution function of the mass ratio of merging binaries. Systems detected with LISA can essentially reproduce very closely the underlying distribution of mass ratios (black dotted line). Therefore, LISA is potentially able to constrain the astrophysical population of merging IMBH binaries for $M_1>10^3\msun$ and any mass ratio. For what concerns ground-based instruments, Voyager is the one that would perform the worst since it will be able to find only near equal-mass binaries ($q>0.3$). Instead, ET and CE will be able to reproduce the distribution of mass ratios in a larger part of the parameter space, $q\gtrsim 0.05$.

The bottom panel of Figure~\ref{fig:muz2} reports the astrophysical distribution of merger redshifts (black dotted line) and the detected distributions by the various observatories considered in this study. LISA, ET, and CE can detect merging systems (in their relevant portion of the $M_1$-$q$ plane) to reproduce the distribution of the merger redshift, with a peak consistent with the characteristics of the astrophysical population. However, Voyager is not be able to reproduce it, being sensitive to merging IMBH binaries at typically lower redshifts.

In Figure~\ref{fig:muz2_multi}, we show the case of multi-band observation. Our criterion for a multi-band detection of a single binary source is that both the observation of its inspiral by LISA plus the observation of its final inspiral or merger or ringdown by a ground-based detector have $\left\langle \rho (z,M_1,q) \right\rangle >8$. It is important to stress that systematic multi-band observations of merging binaries are essentially only possible for systems where one of the components is $\gtrsim 100\msun$. These detections are particularly exciting since they can provide stronger tests of general relativity and cosmology, and can allow tighter constraints on the formation channels of IMBHs \citep{ColpiSesana2017,CarsonYagi2020,GuptaDatta2020,BakerBarausse2022,SeymourYu2022}. We find that LISA + Voyager, LISA + ET, and LISA + CE can observe in multi-band binaries with masses up to $900\msun$, $4000\msun$, and $2000\msun$. Among the different possibilities, LISA + ET is the optimal network since it offers the smallest gap between the two frequency bands. However, the mass ratios of the binaries that can be observed in multi-band is typically $>0.1$. While their observed redshift distributions are still a good representation of the underlying population for LISA + ET and LISA + CE, the detected distribution is still skewed to lower redshifts for LISA + Voyager.

\begin{figure} 
\centering
\includegraphics[scale=0.55]{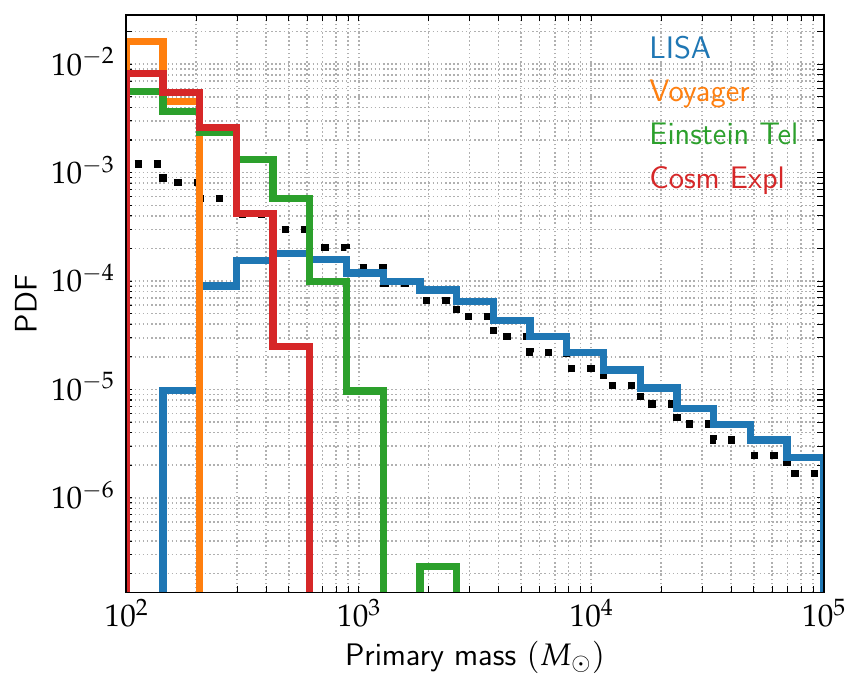}
\includegraphics[scale=0.55]{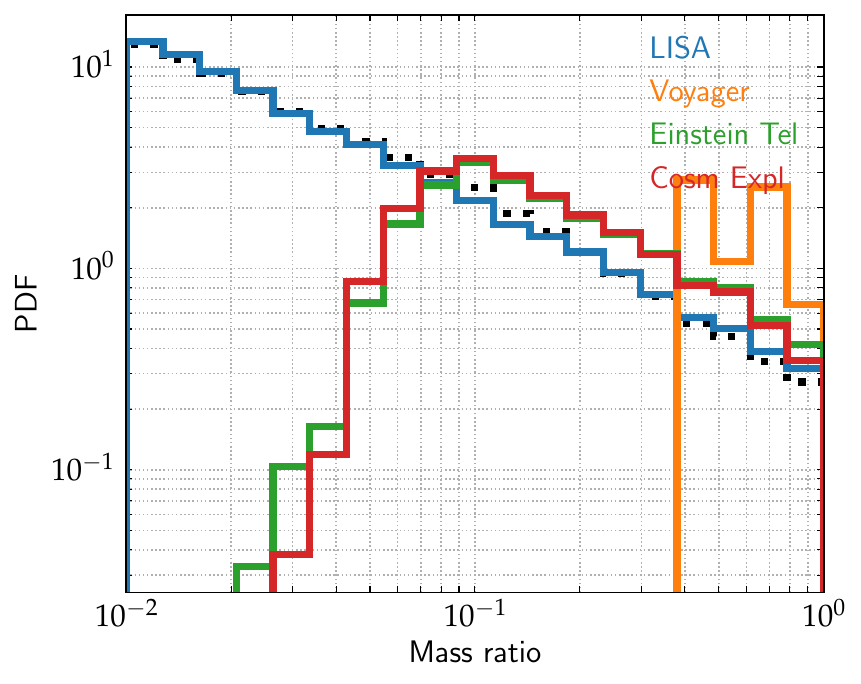}
\includegraphics[scale=0.55]{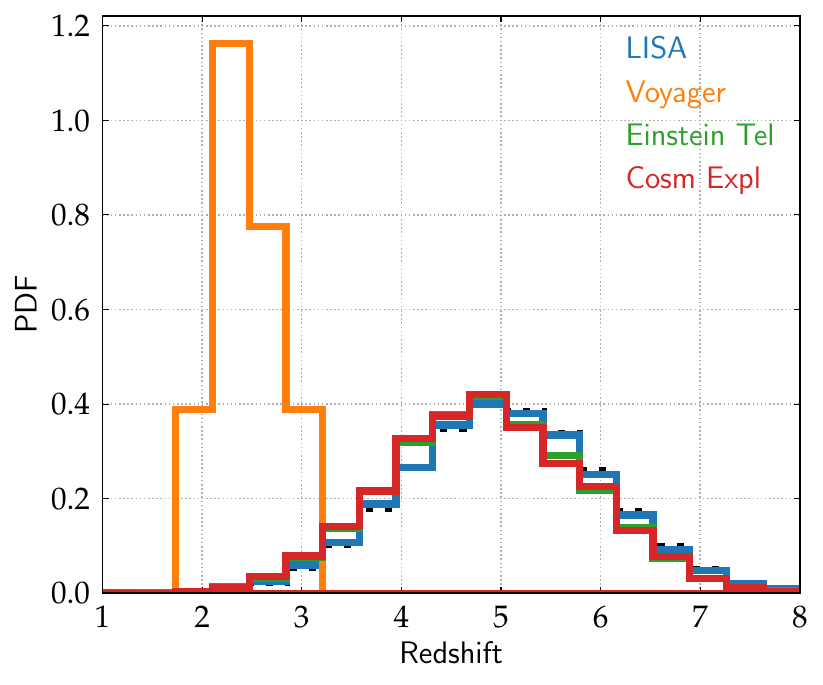}
\caption{Probability distribution function of primary mass (top), mass ratio (center), and redshift (bottom) of merging IMBH binaries detected by LISA (blue), Voyager (orange), Einstein Telescope (green), and Cosmic Explorer (red). The dotted black line represents the underlying sampled population from Eq.~\ref{eqn:nevents} when $\{\mu_z,\sigma_z,\alpha,\beta\}=\{5,1,1,1\}$.}
\label{fig:muz5}
\end{figure}

\begin{figure} 
\centering
\includegraphics[scale=0.55]{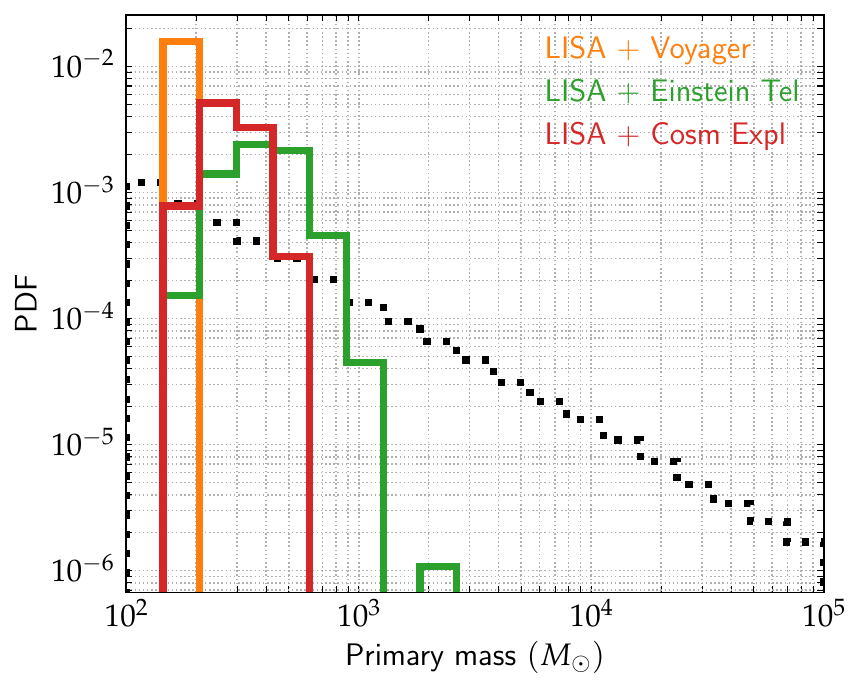}
\includegraphics[scale=0.55]{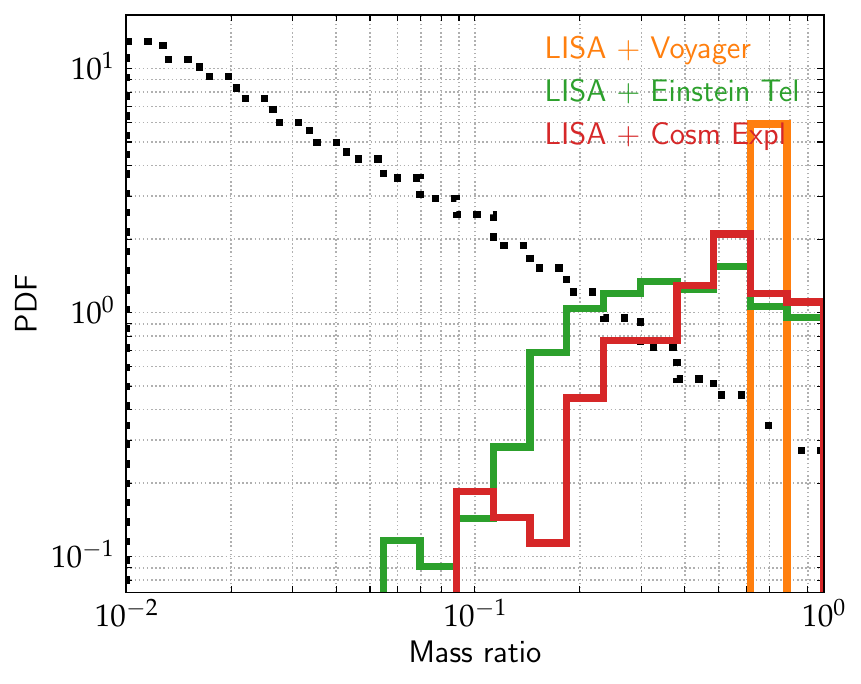}
\includegraphics[scale=0.55]{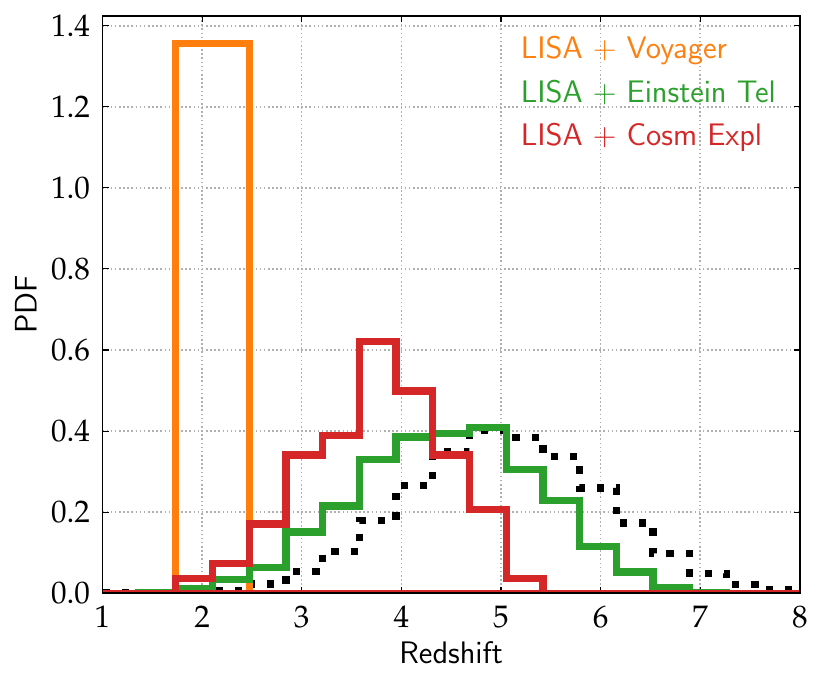}
\caption{Probability distribution function of primary mass (top), mass ratio (center), and redshift (bottom) of merging IMBH binaries detected in multi-band by LISA+Voyager (orange), LISA + Einstein Telescope (green), and LISA + Cosmic Explorer (blue). The dotted black line represents the underlying sampled population from Eq.~\ref{eqn:nevents} when $\{\mu_z,\sigma_z,\alpha,\beta\}=\{5,1,1,1\}$.}
\label{fig:muz5_multi}
\end{figure}

We repeat our analysis in the case where $\{\mu_z,\sigma_z,\alpha,\beta\}=\{5,1,1,1\}$. This choice of parameters may be reminiscent of IMBH mergers from Pop III stars \citep[e.g.,][]{HijikawaKinugawa2022}. Figure~\ref{fig:muz5} shows the probability distribution function of the primary mass (top), mass ratio(center), and redshift (bottom) of merging IMBH binaries detected by LISA, Voyager, ET, and CE. We find that the general trends discussed for the case of $\mu_z=2$ still hold. However, while the population of binaries detected by LISA is not significantly affected, the portion of the parameter space that can be probed with ground-based instruments (and therefore in multi-band, see Figure~\ref{fig:muz5_multi}) is smaller owing to a relative larger merger redshift. Importantly, the highest redshift of a GW source with masses in the IMBH range could test the $\Lambda$CDM paradigm \citep{KoushiappasLoeb2017}.

\section{Discussion and conclusions}
\label{sect:concl}

The direct  measurement of GWs is a powerful tool for surveying the population of BHs across space and time. This is particularly true for IMBHs, which still lack a confirmed detection. However, astrophysical models predict merger rates of IMBHs of $\sim 10^{-3}-10\gpcyr$ in the local Universe, implying several detectable IMBH mergers per year with next-generation ground-based and space-based observatories.

In this paper, we have studied the prospects of using a combination of ground-based (Voyager, Einstein Telescope, Cosmic Explorer) and space-based (LISA) observatories to constrain the cosmic IMBH merger history. We have considered two different scenarios, one representative of IMBH mergers in dynamical environments and one representative of IMBH mergers from Pop III stars. We have found that the binary systems detected by LISA map the underlying astrophysical population for primary masses $\gtrsim 10^3\msun$ and any mass ratio. For primary mass $\lesssim 10^3\msun$ only for $q>0.05$, ET and CE are able to constrain the IMBH merger history, while Voyager could find only about $10\%$ of near equal-mass systems with component masses of $\sim 100\msun$. 

We have also discussed the possibility of multi-band observations. We find that LISA + Voyager, LISA + ET, and LISA + CE can observe in multi-band binaries with masses up to $900\msun$, $4000\msun$, and $2000\msun$, with LISA + ET being the best possible network since it offers the smallest gap between the two frequency bands. However, only binaries with mass ratios $>0.1$ can be detected in multi-band.

We note that we used IMRPhenomD as our approximant for the wavefor, which only includes the dominant harmonic $(l_{\rm GW}, m_{\rm GW}) = (2, 2)$ of the GW signal. However, higher harmonics beyond may contribute in a non-negligible way to the overall SNR for IMBH binaries, in particular for unequal mass-ratios \citep[see e.g.,][]{JaniHealy2016,JaniShoemaker2020}. Nevertheless, current computations of higher harmonics are limited in the parameter space they explore, in particular for what concerns small mass ratios. Therefore, we leave the calculation and inclusion of these higher harmonics to a future work, and stress that our results are a conservative estimation.

The dawn of the GW era of IMBH has just begun. Once operating, next next-generation ground-based and space-based detectors will detect several merging IMBH binaries per year, finally unravelling the mysteries of the most elusive BH population. This may also reveal the growth history of the first quasars.

\section*{Acknowledgements}

G.F.\ acknowledge support from NASA Grant 80NSSC21K1722. A.L.\ was supported in part by the Black Hole Initiative at Harvard University, which is funded by JTF and GBMF grants.

\bibliography{refs}

\begin{thebibliography}{}
\expandafter\ifx\csname natexlab\endcsname\relax\def\natexlab#1{#1}\fi
\providecommand{\url}[1]{\href{#1}{#1}}
\providecommand{\dodoi}[1]{doi:~\href{http://doi.org/#1}{\nolinkurl{#1}}}
\providecommand{\doeprint}[1]{\href{http://ascl.net/#1}{\nolinkurl{http://ascl.net/#1}}}
\providecommand{\doarXiv}[1]{\href{https://arxiv.org/abs/#1}{\nolinkurl{https://arxiv.org/abs/#1}}}

\bibitem[{{Abbott} {et~al.}(2019){Abbott}, {Abbott}, {Abbott}, {Abraham}, \&
  et~al.}]{AbbottAbbott2019}
{Abbott}, B.~P., {Abbott}, R., {Abbott}, T.~D., {Abraham}, S., \& et~al. 2019,
  \prd, 100, 064064, \dodoi{10.1103/PhysRevD.100.064064}

\bibitem[{{Amaro-Seoane} {et~al.}(2017){Amaro-Seoane}, {Audley}, {Babak},
  {Baker}, \& et~al.}]{Amaro-SeoaneAudley2017}
{Amaro-Seoane}, P., {Audley}, H., {Babak}, S., {Baker}, J., \& et~al. 2017,
  arXiv e-prints, arXiv:1702.00786.
\newblock \doarXiv{1702.00786}

\bibitem[{{Antonini} {et~al.}(2019){Antonini}, {Gieles}, \&
  {Gualandris}}]{antonini2019}
{Antonini}, F., {Gieles}, M., \& {Gualandris}, A. 2019, \mnras, 486, 5008,
  \dodoi{10.1093/mnras/stz1149}

\bibitem[{{Arca Sedda} {et~al.}(2021){Arca Sedda}, {Amaro Seoane}, \&
  {Chen}}]{ArcaSeddaAmaroSeoane2021}
{Arca Sedda}, M., {Amaro Seoane}, P., \& {Chen}, X. 2021, \aap, 652, A54,
  \dodoi{10.1051/0004-6361/202037785}

\bibitem[{{Arca-Sedda} \&
  {Capuzzo-Dolcetta}(2019)}]{Arca-SeddaCapuzzo-Dolcetta2019}
{Arca-Sedda}, M., \& {Capuzzo-Dolcetta}, R. 2019, \mnras, 483, 152,
  \dodoi{10.1093/mnras/sty3096}

\bibitem[{{Arca-Sedda} \& {Gualandris}(2018)}]{agu18}
{Arca-Sedda}, M., \& {Gualandris}, A. 2018, \mnras, 477, 4423,
  \dodoi{10.1093/mnras/sty922}

\bibitem[{{Baker} {et~al.}(2022){Baker}, {Barausse}, {Chen}, {de Rham}, \&
  et~al.}]{BakerBarausse2022}
{Baker}, T., {Barausse}, E., {Chen}, A., {de Rham}, C., \& et~al. 2022, arXiv
  e-prints, arXiv:2209.14398.
\newblock \doarXiv{2209.14398}

\bibitem[{{Baldassare} {et~al.}(2015){Baldassare}, {Reines}, {Gallo}, \&
  {Greene}}]{BaldassareReines2015}
{Baldassare}, V.~F., {Reines}, A.~E., {Gallo}, E., \& {Greene}, J.~E. 2015,
  \apjl, 809, L14, \dodoi{10.1088/2041-8205/809/1/L14}

\bibitem[{{Begelman} {et~al.}(2006){Begelman}, {Volonteri}, \&
  {Rees}}]{begelm2006}
{Begelman}, M.~C., {Volonteri}, M., \& {Rees}, M.~J. 2006, \mnras, 370, 289,
  \dodoi{10.1111/j.1365-2966.2006.10467.x}

\bibitem[{{Bromm} \& {Larson}(2004)}]{bromm2004}
{Bromm}, V., \& {Larson}, R.~B. 2004, \araa, 42, 79,
  \dodoi{10.1146/annurev.astro.42.053102.134034}

\bibitem[{{Bromm} \& {Loeb}(2003)}]{bromm2003}
{Bromm}, V., \& {Loeb}, A. 2003, \apj, 596, 34, \dodoi{10.1086/377529}

\bibitem[{{Carson} \& {Yagi}(2020)}]{CarsonYagi2020}
{Carson}, Z., \& {Yagi}, K. 2020, Classical and Quantum Gravity, 37, 02LT01,
  \dodoi{10.1088/1361-6382/ab5c9a}

\bibitem[{{Chilingarian} {et~al.}(2018){Chilingarian}, {Katkov}, {Zolotukhin},
  {Grishin}, {Beletsky}, {Boutsia}, \& {Osip}}]{chili2018}
{Chilingarian}, I.~V., {Katkov}, I.~Y., {Zolotukhin}, I.~Y., {et~al.} 2018,
  \apj, 863, 1, \dodoi{10.3847/1538-4357/aad184}

\bibitem[{{Colpi} \& {Sesana}(2017)}]{ColpiSesana2017}
{Colpi}, M., \& {Sesana}, A. 2017, in An Overview of Gravitational Waves:
  Theory, 43--140, \dodoi{10.1142/9789813141766_0002}

\bibitem[{{Di Carlo} {et~al.}(2021){Di Carlo}, {Mapelli}, {Pasquato},
  {Rastello}, \& et~al.}]{DiCarloMapelli2021}
{Di Carlo}, U.~N., {Mapelli}, M., {Pasquato}, M., {Rastello}, S., \& et~al.
  2021, \mnras, \dodoi{10.1093/mnras/stab2390}

\bibitem[{{Di Matteo} {et~al.}(2022){Di Matteo}, {Ni}, {Chen}, {Croft}, \&
  et~al.}]{DiMatteoNi2022}
{Di Matteo}, T., {Ni}, Y., {Chen}, N., {Croft}, R., \& et~al. 2022, arXiv
  e-prints, arXiv:2210.14960.
\newblock \doarXiv{2210.14960}

\bibitem[{{Fragione}(2022)}]{Fragione2022}
{Fragione}, G. 2022, \apj, 939, 97, \dodoi{10.3847/1538-4357/ac98b6}

\bibitem[{Fragione(2023)}]{imbhistory}
Fragione, G. 2023, {imbhistory}, v1.01,  Zenodo, \dodoi{10.5281/zenodo.7530024}

\bibitem[{{Fragione} {et~al.}(2018{\natexlab{a}}){Fragione}, {Ginsburg}, \&
  {Kocsis}}]{fragk18}
{Fragione}, G., {Ginsburg}, I., \& {Kocsis}, B. 2018{\natexlab{a}}, \apj, 856,
  92, \dodoi{10.3847/1538-4357/aab368}

\bibitem[{{Fragione} {et~al.}(2022){Fragione}, {Kocsis}, {Rasio}, \&
  {Silk}}]{FragioneKocsis2022}
{Fragione}, G., {Kocsis}, B., {Rasio}, F.~A., \& {Silk}, J. 2022, \apj, 927,
  231, \dodoi{10.3847/1538-4357/ac5026}

\bibitem[{{Fragione} \& {Leigh}(2018)}]{fragl2018b}
{Fragione}, G., \& {Leigh}, N. 2018, \mnras, 480, 5160,
  \dodoi{10.1093/mnras/sty2233}

\bibitem[{{Fragione} {et~al.}(2018{\natexlab{b}}){Fragione}, {Leigh},
  {Ginsburg}, \& {Kocsis}}]{fragleiginkoc18}
{Fragione}, G., {Leigh}, N. W.~C., {Ginsburg}, I., \& {Kocsis}, B.
  2018{\natexlab{b}}, \apj, 867, 119, \dodoi{10.3847/1538-4357/aae486}

\bibitem[{{Gair} {et~al.}(2011){Gair}, {Mandel}, {Miller}, \&
  {Volonteri}}]{gair2011}
{Gair}, J.~R., {Mandel}, I., {Miller}, M.~C., \& {Volonteri}, M. 2011, General
  Relativity and Gravitation, 43, 485, \dodoi{10.1007/s10714-010-1104-3}

\bibitem[{{Giersz} {et~al.}(2015){Giersz}, {Leigh}, {Hypki}, {L\"{u}tzgendorf},
  \& {Askar}}]{gie15}
{Giersz}, M., {Leigh}, N.~W., {Hypki}, A., {L\"{u}tzgendorf}, N., \& {Askar},
  A. 2015, \mnras, 454, 3150, \dodoi{10.1093/mnras/stv2162}

\bibitem[{{Gonz{\'a}lez} {et~al.}(2021){Gonz{\'a}lez}, {Kremer}, {Chatterjee},
  {Fragione}, \& et~al.}]{GonzalezKremer2021}
{Gonz{\'a}lez}, E., {Kremer}, K., {Chatterjee}, S., {Fragione}, G., \& et~al.
  2021, \apjl, 908, L29, \dodoi{10.3847/2041-8213/abdf5b}

\bibitem[{{Greene} {et~al.}(2020){Greene}, {Strader}, \&
  {Ho}}]{GreeneStrader2020}
{Greene}, J.~E., {Strader}, J., \& {Ho}, L.~C. 2020, \araa, 58, 257,
  \dodoi{10.1146/annurev-astro-032620-021835}

\bibitem[{{Gupta} {et~al.}(2020){Gupta}, {Datta}, {Kastha}, {Borhanian}, \&
  et~al.}]{GuptaDatta2020}
{Gupta}, A., {Datta}, S., {Kastha}, S., {Borhanian}, S., \& et~al. 2020, \prl,
  125, 201101, \dodoi{10.1103/PhysRevLett.125.201101}

\bibitem[{{G{\"u}rkan} {et~al.}(2004){G{\"u}rkan}, {Freitag}, \&
  {Rasio}}]{gurk2004}
{G{\"u}rkan}, M.~A., {Freitag}, M., \& {Rasio}, F.~A. 2004, \apj, 604, 632,
  \dodoi{10.1086/381968}

\bibitem[{{Hijikawa} {et~al.}(2022){Hijikawa}, {Kinugawa}, {Tanikawa},
  {Yoshida}, \& et~al.}]{HijikawaKinugawa2022}
{Hijikawa}, K., {Kinugawa}, T., {Tanikawa}, A., {Yoshida}, T., \& et~al. 2022,
  arXiv e-prints, arXiv:2211.07496.
\newblock \doarXiv{2211.07496}

\bibitem[{{Husa} {et~al.}(2016){Husa}, {Khan}, {Hannam}, {P{\"u}rrer}, \&
  et~al.}]{HusaKhan2016}
{Husa}, S., {Khan}, S., {Hannam}, M., {P{\"u}rrer}, M., \& et~al. 2016, \prd,
  93, 044006, \dodoi{10.1103/PhysRevD.93.044006}

\bibitem[{{Jani} {et~al.}(2016){Jani}, {Healy}, {Clark}, {London}, \&
  et~al.}]{JaniHealy2016}
{Jani}, K., {Healy}, J., {Clark}, J.~A., {London}, L., \& et~al. 2016,
  Classical and Quantum Gravity, 33, 204001,
  \dodoi{10.1088/0264-9381/33/20/204001}

\bibitem[{{Jani} {et~al.}(2020){Jani}, {Shoemaker}, \&
  {Cutler}}]{JaniShoemaker2020}
{Jani}, K., {Shoemaker}, D., \& {Cutler}, C. 2020, Nature Astronomy, 4, 260,
  \dodoi{10.1038/s41550-019-0932-7}

\bibitem[{{Kaaret} {et~al.}(2017){Kaaret}, {Feng}, \&
  {Roberts}}]{kaaret2017ARA&A..55..303K}
{Kaaret}, P., {Feng}, H., \& {Roberts}, T.~P. 2017, \araa, 55, 303,
  \dodoi{10.1146/annurev-astro-091916-055259}

\bibitem[{{Kocsis} {et~al.}(2011){Kocsis}, {Yunes}, \& {Loeb}}]{koc11}
{Kocsis}, B., {Yunes}, N., \& {Loeb}, A. 2011, \prd, 84, 024032,
  \dodoi{10.1103/PhysRevD.84.024032}

\bibitem[{{Koushiappas} \& {Loeb}(2017)}]{KoushiappasLoeb2017}
{Koushiappas}, S.~M., \& {Loeb}, A. 2017, \prl, 119, 221104,
  \dodoi{10.1103/PhysRevLett.119.221104}

\bibitem[{{LIGO Scientific Collaboration}(2018)}]{LIGOVoy2018}
{LIGO Scientific Collaboration}. 2018, {Instrument Science White Paper 2018}

\bibitem[{{Lin} {et~al.}(2018){Lin}, {Strader}, {Carrasco}, {Page},
  {Romanowsky}, {Homan}, {Irwin}, {Remillard}, {Godet}, {Webb}, {Baumgardt},
  {Wijnands}, {Barret}, {Duc}, {Brodie}, \& {Gwyn}}]{lin2018}
{Lin}, D., {Strader}, J., {Carrasco}, E.~R., {et~al.} 2018, Nature Astronomy,
  2, 656, \dodoi{10.1038/s41550-018-0493-1}

\bibitem[{{MacLeod} {et~al.}(2016){MacLeod}, {Guillochon}, {Ramirez-Ruiz},
  {Kasen}, \& et~al.}]{MacLeodGuillochon2016}
{MacLeod}, M., {Guillochon}, J., {Ramirez-Ruiz}, E., {Kasen}, D., \& et~al.
  2016, \apj, 819, 3, \dodoi{10.3847/0004-637X/819/1/3}

\bibitem[{{Madau} \& {Rees}(2001)}]{madau2001}
{Madau}, P., \& {Rees}, M.~J. 2001, \apjl, 551, L27, \dodoi{10.1086/319848}

\bibitem[{{Mandel} {et~al.}(2008){Mandel}, {Brown}, {Gair}, \&
  {Miller}}]{MandelBrown2008}
{Mandel}, I., {Brown}, D.~A., {Gair}, J.~R., \& {Miller}, M.~C. 2008, \apj,
  681, 1431, \dodoi{10.1086/588246}

\bibitem[{{Miller}(2002)}]{Miller2002}
{Miller}, M.~C. 2002, \apj, 581, 438, \dodoi{10.1086/344156}

\bibitem[{{Miller} \& {Hamilton}(2002)}]{mil02b}
{Miller}, M.~C., \& {Hamilton}, D.~P. 2002, \mnras, 330, 232,
  \dodoi{10.1046/j.1365-8711.2002.05112.x}

\bibitem[{{Natarajan}(2021)}]{Natarajan2021}
{Natarajan}, P. 2021, \mnras, 501, 1413, \dodoi{10.1093/mnras/staa3724}

\bibitem[{{Nitz} {et~al.}(2019){Nitz}, {Harry}, {Brown}, {Biwer}, \&
  et~al.}]{NitzHarry2019}
{Nitz}, A., {Harry}, I., {Brown}, D., {Biwer}, C.~M., \& et~al. 2019,
  {gwastro/pycbc: PyCBC Release v1.14.4}, v1.14.4, Zenodo,  Zenodo,
  \dodoi{10.5281/zenodo.3546372}

\bibitem[{{Pechetti} {et~al.}(2022){Pechetti}, {Seth}, {Kamann}, {Caldwell}, \&
  et~al.}]{PechettiSeth2022}
{Pechetti}, R., {Seth}, A., {Kamann}, S., {Caldwell}, N., \& et~al. 2022, \apj,
  924, 48, \dodoi{10.3847/1538-4357/ac339f}

\bibitem[{{Peng} {et~al.}(2019){Peng}, {Yang}, {Shen}, {Wang}, {Zou}, \&
  {Zhang}}]{peng2019}
{Peng}, Z.-K., {Yang}, Y.-S., {Shen}, R.-F., {et~al.} 2019, \apjl, 884, L34,
  \dodoi{10.3847/2041-8213/ab481b}

\bibitem[{{Planck Collaboration} {et~al.}(2016){Planck Collaboration}, {Ade},
  {Aghanim}, {Arnaud}, \& et~al.}]{PlanckCollaborationAde2016}
{Planck Collaboration}, {Ade}, P.~A.~R., {Aghanim}, N., {Arnaud}, M., \& et~al.
  2016, \aap, 594, A13, \dodoi{10.1051/0004-6361/201525830}

\bibitem[{{Portegies Zwart} \& {McMillan}(2002)}]{por02}
{Portegies Zwart}, S.~F., \& {McMillan}, S.~L.~W. 2002, \apj, 576, 899,
  \dodoi{10.1086/341798}

\bibitem[{{Punturo} {et~al.}(2010){Punturo}, {Abernathy}, {Acernese}, {Allen},
  \& et~al.}]{PunturoAbernathy2010}
{Punturo}, M., {Abernathy}, M., {Acernese}, F., {Allen}, B., \& et~al. 2010,
  Classical and Quantum Gravity, 27, 194002,
  \dodoi{10.1088/0264-9381/27/19/194002}

\bibitem[{{Rasskazov} {et~al.}(2020){Rasskazov}, {Fragione}, \&
  {Kocsis}}]{RasskazovFragione2020}
{Rasskazov}, A., {Fragione}, G., \& {Kocsis}, B. 2020, \apj, 899, 149,
  \dodoi{10.3847/1538-4357/aba2f4}

\bibitem[{{Reitze} {et~al.}(2019){Reitze}, {Adhikari}, {Ballmer}, {Barish}, \&
  et~al.}]{ReitzeAdhikari2019}
{Reitze}, D., {Adhikari}, R.~X., {Ballmer}, S., {Barish}, B., \& et~al. 2019,
  in Bulletin of the American Astronomical Society, Vol.~51, 35.
\newblock \doarXiv{1907.04833}

\bibitem[{{Rizzuto} {et~al.}(2022){Rizzuto}, {Naab}, {Rantala}, {Johansson}, \&
  et~al.}]{RizzutoNaab2022}
{Rizzuto}, F.~P., {Naab}, T., {Rantala}, A., {Johansson}, P.~H., \& et~al.
  2022, arXiv e-prints, arXiv:2211.13320.
\newblock \doarXiv{2211.13320}

\bibitem[{Robson {et~al.}(2019)Robson, Cornish, \& Liu}]{robson2019}
Robson, T., Cornish, N.~J., \& Liu, C. 2019, Classical and Quantum Gravity, 36,
  105011, \dodoi{10.1088/1361-6382/ab1101}

\bibitem[{{Rose} {et~al.}(2022){Rose}, {Naoz}, {Sari}, \&
  {Linial}}]{RoseNaoz2022}
{Rose}, S.~C., {Naoz}, S., {Sari}, R., \& {Linial}, I. 2022, \apjl, 929, L22,
  \dodoi{10.3847/2041-8213/ac6426}

\bibitem[{{Rosswog} {et~al.}(2009){Rosswog}, {Ramirez-Ruiz}, \&
  {Hix}}]{RosswogRamirez-Ruiz2009}
{Rosswog}, S., {Ramirez-Ruiz}, E., \& {Hix}, W.~R. 2009, \apj, 695, 404,
  \dodoi{10.1088/0004-637X/695/1/404}

\bibitem[{{Seymour} {et~al.}(2022){Seymour}, {Yu}, \& {Chen}}]{SeymourYu2022}
{Seymour}, B.~C., {Yu}, H., \& {Chen}, Y. 2022, arXiv e-prints,
  arXiv:2208.01668.
\newblock \doarXiv{2208.01668}

\bibitem[{{Shen}(2019)}]{shen2019}
{Shen}, R.-F. 2019, \apjl, 871, L17, \dodoi{10.3847/2041-8213/aafc64}

\bibitem[{{Silk}(2017)}]{Silk2017}
{Silk}, J. 2017, \apjl, 839, L13, \dodoi{10.3847/2041-8213/aa67da}

\bibitem[{{Stone} {et~al.}(2017){Stone}, {K{\"u}pper}, \&
  {Ostriker}}]{StoneKupper2017}
{Stone}, N.~C., {K{\"u}pper}, A. H.~W., \& {Ostriker}, J.~P. 2017, \mnras, 467,
  4180, \dodoi{10.1093/mnras/stx097}

\end{thebibliography}

\end{document}